\documentclass[12pt]{article}   
\title{Running couplings and triviality of field theories on 
non-commutative spaces}  

\author{Emil T. Akhmedov\footnote{Permanent Address:
Institute of Theoretical and Experimental Physics,
B. Cheremushkinskaya 25, 117259 Moscow, Russia}, 
Philip DeBoer and  Gordon W. Semenoff\\~~\\
Department of Physics and Astronomy, \\
University of British Columbia, \\
Vancouver, British Columbia, Canada V6T 1Z1.}

\begin{document}           

\maketitle                 

\abstract{We examine the issue of renormalizability of
asymtotically free field theories on non-commutative spaces.  As an
example, we solve the non-commutative O(N) invariant Gross-Neveu model
at large N.  On commutative space this is a renormalizable model with
non-trivial interactions.  On the noncommutative space, if we take the
translation invariant ground state, we find that the model is
non-renormalizable.  Removing the ultraviolet cutoff yields a
trivial non-interacting theory.}

\newpage

\noindent
{\bf 1.} Motivated by the fact that they arise as the low energy
limits of string theories with antisymmetric tensor backgrounds
\cite{Ho:1997jr,Connes:1998cr,Chu:2000gi,Seiberg:1999vs,
Lee:2000kj,Laidlaw:2000kb}, 
field theories on non-commutative spaces have recently received a
great deal of attention.  On non-commutative $R^d$, the coordinates
have the algebra $
\left[ x^\mu, x^\nu \right] = -i\theta^{\mu\nu}
$, where $\theta$ is an antisymmetric matrix.  This noncommutativity
can also be implemented by replacing the usual product for multiplying
functions by the associative, non-commutative and non-local $*$-product,
\begin{equation}
f(x)*g(x)\equiv \lim_{y\rightarrow x}
\exp\left( -\frac{i}{2}\theta^{\mu\nu}\frac{\partial}
{\partial x^\mu}\frac{\partial}{\partial y^\nu}\right) f(x)g(y)
\label{noncom}
\end{equation}
When non-commutative field theories are obtained as limits of string
theory, the massive string excitations decouple.  For this reason the
non-commutative field theories are believed to be unitary and
renormalizable.  In fact, for some theories, unitarity has been
denomstrated explicitly at one-loop order \cite{Gomis:2000zz}.  This
raises the interesting possibility that there exist consistent,
unitary, and non-trivial nonlocal field theories.  In this Letter, we
examine this issue in the non-perturbative context of a solvable
model.

There are several features of non-commutative field theories which
distinguish them from their commutative analogs.  One already occurs
in scalar field theory.  The non-commutativity affects the spectrum
and interactions of the theory at low energy scales, below the
momentum scale $1/\sqrt{|\theta|}$ set by the dimensional parameter
$\theta_{\mu\nu}$ and in fact below the mass scales of the particles
already in the model.  This has been associated with the phenomenon of
ultraviolet/infrared (UV/IR) duality familiar from the behavior of
D-branes in string theory \cite{Minwalla:1999px,VanRaamsdonk:2000rr}.

In non-commutative field theory, there is a sense in which
perturbative contributions to a given diagram can be divided into
planar and non-planar graphs \cite{Filk:1996dm}.  The Feynman
integrands of planar graphs are as they were in the commutative case.
The integrands of non-planar graphs are modified by phases containing
external and internal loop momenta.  The presence of these phases
improves the high-momentum behavior of Feynman integrals.  The most
dramatic effect occurs in diagrams which are ultraviolet divergent.
Planar diagrams diverge and must be defined using a high momentum
cutoff, $\Lambda$.  The non-planar diagrams generally converge, the
ultraviolet cutoff being replaced by 
$\Lambda_{\rm eff}(p)=\Lambda/
\sqrt{1+\Lambda^2(\theta p)^2}$. For any non-zero momentum, 
this effective cutoff has a finite limit, $\Lambda_{\rm eff}\sim
1/|\theta p|$, as $\Lambda\rightarrow\infty$.

For example, at one loop order in 4-dimensional $\phi^4$-theory, the
radiative correction to the scalar self-energy in the commutative
version is a quadratically divergent constant.  In the non-commutative
theory, there are two contributions, a planar and non-planar one.  The
non-planar one turns out to be a function of external momentum.  At
small momentum, \cite{Minwalla:1999px,VanRaamsdonk:2000rr}

\begin{equation}
\Gamma^{(2)}(p)=
\frac{g^2}{48\pi^2}\left( \Lambda^2-m^2\ln\frac{\Lambda^2}{m^2}\right)+
\frac{g^2}{96\pi^2}\left( \Lambda_{\rm eff}^2(p)-m^2\ln\frac{\Lambda_{\rm eff}^2(p)}{m^2}\right)+
\ldots
\label{twopt}
\end{equation}
Here, $m$ is the scalar field mass and $g^2$ is the dimensionless
$\phi^4$ coupling constant.
The second term has a pole at very low momenta.  It also has a logarithmic
cut singularity at small momenta.
These 
have been argued to arise from new degrees
of freedom with exotic dispersion relations or perhaps propagating in higher
dimensions \cite{VanRaamsdonk:2000rr}.
They have also been argued to lead to exotic translation 
non-invariant ``striped'' 
phases of scalar field theory \cite{Gubser:2000cd}.

Another place where ultraviolet divergences typically 
occur and have interesting effects is in the renormalization of dimensionless
coupling constants.  
The leading corrections to the coupling constant in 4-dimensional
$\phi^4$-theory are logarithmically divergent.  
The small momentum limit of the 4-point function
was computed in \cite{Minwalla:1999px} as

\begin{eqnarray}
\Gamma^{(4)}(p,q,r,s)=g^2-\frac{g^2}{2\cdot 2^5\pi^2}\left\{
2\ln\frac{\Lambda^2}{m^2}+\ln\frac{1}{m^2(\theta p)^2}+\ln\frac{1}{m^2(\theta q)^2}
+
\right.
\nonumber \\
\left. +\ln\frac{1}{m^2(\theta r)^2} 
+\ln\frac{1}{m^2(\theta s)^2}
+\ln\frac{1}{m^2(\theta(q+r) )^2}
+\right.
\nonumber \\
\left.
+\ln\frac{1}{m^2(\theta(q+s))^2}
+\ln\frac{1}{m^2(\theta(r+s))^2} \right\}+\ldots
\label{fourpt}
\end{eqnarray}
($p+q+r+s=0$) The first contribution is from planar diagrams and the
others are from non-planar diagrams and they depend explicitly on the
parameter $\theta$.  What is remarkable about (\ref{fourpt}) is that,
in spite of the mass gap, the effective coupling constant is
logarithmically singular at small momentum, similar to its large
momentum limit but with $p^2/\Lambda^2$ replaced by $m^2(\theta p)^2$.
If we were to sum the leading logarithmically singular diagrams to all
orders, we would obtain a coupling constant which runs at small
momentum scales.  This is distinct from the behavior of commutative
field theory where the running of coupling constants is cutoff by mass
scales.  For example, in quantum electrodynamics which, like
$\phi^4$-theory is infrared free, the coupling constant runs at
energies much larger than the electron mass, but as the energy is
lowered, it freezes at the value $e^2/4\pi\sim 1/137$ and is the same
at all lower energy scales.  In a noncommutative theory, it appears
that masses do not cutoff the running of coupling constants.  We note
that this is a non-perturbative issue, which occurs in addition to the
perturbative renormalizability of non-commutative field theories which
has recently been examined in detail
\cite{Chepelev:2000tt,Chepelev:2000hm,Sheikh}.

Of particular interest are asymptotically free field theories.  In
this Letter, we shall examine a simple asymptotically free field
theory, the $O(N)$ Gross-Neveu model \cite{Gross:1974jv} of
interacting fermions in two spacetime dimensions.  In this model mass
renormalization is protected by symmetry, so the behavior analogous to
(\ref{twopt}) is absent.  However, it does have logarithmic coupling
constant renormalization and we are able to examine its effect in the
large $N$ limit which sums all orders in perturbation theory.  We
shall find that the theory is not renormalizable.  As a consequence,
if we require that the dynamically generated fermion mass is finite
\footnote{The cutoff dependence of the fermion mass comes from the
gap equation.  If a different dependence were chosen, there would be
tachions in the spectrum of the theory.}, the effective four-fermi
coupling is still cutoff dependent and goes to zero as the cutoff is
removed.  It also exhibits an interesting UV/IR duality.  In the large
momentum limit it has the momentum dependence $$
\lambda_{eff}(p) = \frac{8\pi}{\ln{\frac{\Lambda^2 p^2}{M^4}}},
$$ where $M$ is the dynamically generated fermion mass.  
There is a low-energy mirror of this behavior in the infrared, 
for momenta in the range $ 1/(\theta\Lambda)^2<<p^2<<1/(\theta
M)^2$, $$
\lambda_{eff}(p) = \frac{8\pi}{ \ln\left(\Lambda^2 (\theta p)^2\right)}
$$ 
This is an example of a field theory that is perfectly
renormalizable and nontrivial on a commutative space and is not
renormalizable and has trivial correlators in the infinite cutoff
limit on a non-commutative space. It can only be non-trivial if the
coupling is kept finite.  This is in line with suggestions that
non-commutative field theories are only well-defined when there is a
finite cutoff in both the UV and IR \cite{Ambjorn:2000nb}.

There is a renormalizable double-scaling limit of the theory that can
be taken and in which the cut-off dependence is removed.  In this
limit, $\theta\rightarrow 0$ and $\Lambda\rightarrow \infty$ holding
$\theta\Lambda$ fixed.  Space-time is non-commutative
only on distance scales of order the UV cutoff. We shall find that,
nevertheless, the theory in this limit exhibits a behavior which is
quite different from its commutative analog.

\vskip 0.2 cm
\noindent
{\bf 2.} The Euclidean action of the non-commutative Gross-Neveu model is

\begin{equation}
S[\psi]=-\int d^2x \left\{ \frac{1}{2}\sum_{j=1}^N
\bar\psi^j\gamma\cdot\partial\psi^j
+\frac{\lambda}{8N}
\sum_{ij=1}^N\bar\psi^i*\psi^i*\bar\psi^j*\psi^j\right\}
\label{gn}
\end{equation}
Here $\psi^i$ are N 2-component Majorana fermions.  They obey the
constraint $\psi= C\psi^*$ with $C$ the charge conjugation matrix.  We
use Majorana, rather than Dirac fermions because in the latter case,
corrections to the four fermion interaction are not sensitive to the
non-commutativity in the leading order in 1/N.

The kinetic term in (\ref{gn}) has $O(N)_L\times O(N)_R$ chiral
symmetry.  The interaction term breaks this to a diagonal $O(N)$.
There is also a discrete chiral symmetry, $\psi\rightarrow
\gamma^5\psi$ with $\gamma^5=i\gamma^1\gamma^2$.  The condensate
$\left<\bar\psi\psi\right>$ is an order parameter for breaking of this
symmetry.  If it is non-zero, the fermions are massive.  All products
in (\ref{gn}) are $*$ products, as defined in (\ref{noncom}).  When we
set total derivative terms in the action (\ref{gn}) to zero, in each
term, one of the $*$-products is always equal
to an ordinary product.  For this reason only the ordinary product
occurs in quadratic terms.  The quartic term can be written as
$\int(\bar\psi*\psi)^2$.  We shall use this fact later when we
introduce an auxiliary field.  Note that in two dimensions
non-commutativity does not break Lorenz invariance,  
$\theta_{\mu\nu} = \theta \cdot
\epsilon_{\mu\nu}$.  Also,
in Minkowski space, it would not be possible to set $\theta^{0i}$ to
zero, so a 1+1-dimensional noncommutative theory may not be a
well-defined Hamiltonian system.  Here we take the philosophy that
(\ref{gn}) does define a statistical model where correlators can be
computed and where issues such as renormability, which do occur in
other more physical models can be addresed.

This model is explicitly solvable in the infinite $N$ limit.  The
commutative version is asymptotically free.  The effective
four-fermion coupling decreases with increasing momentum transfer and
increases with decreasing momentum, running to strong coupling in the
infrared.  This running is cut off by spontaneous generation of the
fermion mass.  The result is a non-trivial, interacting field theory
with a dynamically generated mass scale.  This is prototypical of some
of the behavior which is thought to occur in other asymptotically free
theories such as four dimensional Yang-Mills theory.

In order to solve the large $N$ limit of the model (\ref{gn}), it is
convenient to introduce an auxiliary field so that the action is

\begin{equation}
S[\psi,\phi]=\int d^2x \left\{ -\frac{1}{2}\sum_{j=1}^N
\bar\psi^j(\gamma\cdot\partial+*\phi*)\psi^j
+\frac{N}{2\lambda}\phi^2\right\}
\label{gn1}
\end{equation}
The original action (\ref{gn}) is re-obtained by integrating out
$\phi$ in the partition function, $Z=\int [d\psi
d\phi]\exp\left(-S[\psi,\phi]\right)$.  Instead, we integrate out
${\psi}$ to get the non-local scalar field theory with action

\begin{equation}
S[\phi]= -\frac{N}{2}{\rm Tr}\ln \left( \gamma\cdot\partial + *\phi*\right)
+\int\frac{N}{2\lambda}\phi^2
\label{fermact}
\end{equation}
where $*\phi*$ denotes multiplication using the $*$-product.  In the
large $N$ limit, the remaining functional integral can be evaluated by
saddle point approximation.  For this, we must find a minimum of
(\ref{fermact}) as a functional of $\phi$.  We will restrict our
search for minima to those which give translation invariant ground
states, i.e. to where the function $\phi$ which minimizes
(\ref{fermact}) is a constant.  At this point, we do not know whether
there are translation non-invariant solutions which are have smaller
action than the translation invariant solution that we find.  We will
not address this issue in this Letter. Since we find no tachyons in
the spectrum, the solution which we consider is at least a local
minumum.  If $\phi=M$ is a constant, we can readily
evaluate (\ref{fermact}).  Divided by the space-time volume, it is the
effective potential
\begin{equation}
V_{\rm eff}=-\frac{N}{8\pi}\left( M^2\ln\frac{\Lambda^2}{M^2}+M^2\right)
+\frac{N}{2{\lambda}} M^2
\label{veff}
\end{equation} 
which must be minimized in order to find the physical value of $M$.  It
always has a minimum for non-zero $M$ which occurs when
\begin{equation}
\frac{1}{\lambda}=\frac{1}{4\pi}\ln\frac{\Lambda^2}{M^2}
\label{gap}
\end{equation}
This equation is dimensional transmutation: the bare dimensionless
coupling $\lambda$ and UV cutoff are traded for a dimensional parameter,
$M$, the dynamically generated fermion mass.

The effective four-fermion coupling is determined by the quadratic
fluctuations of $\phi$.  Consider $\phi=M+\delta\phi$ in
(\ref{fermact}).  Since $\delta\phi$ is not a constant, we must be
careful to take into account the $*$-product in the determinant. 
The determinant is defined by the expression

\begin{equation}
-\frac{1}{2}\ln{\rm Tr}\left( \gamma\cdot\partial + *\phi*\right)
=\sum_{n=1}^\infty\frac{1}{n!}\int \delta\phi(x_1)
\ldots \delta\phi(x_n)\tau(x_1,\ldots,x_n)
\end{equation}
where

\begin{equation}
\tau(x_1,\ldots,x_n)=-\left(\frac{1}{2}\right)^n
\left< \bar\psi(x_1)*\psi(x_1)\ldots\bar\psi(x_n)
*\psi(x_n)\right>^{\rm conn.}_0
\end{equation}
The expectation values are taken with respect to free fermions with
mass $M$.  Of particular interest is the quadratic term with 
$\tau(x_1,x_2)=
\int \frac{ d^2q}{(2\pi)^2}\tau(q) e^{iq\cdot(x_1-x_2)}$.  
It gets a contribution from a planar and a non-planar diagram and thus
depends on the non-commutativity parameter.  The planar diagram
contributes

\begin{equation}
\tau_1(q)=- \frac{1}{4\pi}\left( \ln\frac{\Lambda e^{1-\gamma}}{M} 
-\frac{ \sqrt{1+\frac{q^2}{4M^2}} }{\frac{q}{2M}}\ln\left(\sqrt{ 1+\frac{q^2}{4M^2}}+\frac{q}{2M}\right) 
\right)\end{equation}
where $\gamma$ is Euler's constant.  Here, we have used the same regularization
as in \cite{Minwalla:1999px} and \cite{VanRaamsdonk:2000rr}.
The non-planar contribution is 

\begin{eqnarray}
\tau_2(q)= - \frac{1}{4\pi}
K_0\left(2M
\sqrt{[\theta^2q^2/4+1/\Lambda^2]}\right) +~~~~~~~~~~~~~~~~~~~~~~~~~~~
\nonumber \\
+ \frac{1}{4\pi}\left(M^2 + \frac{q^2}{4}\right) \int^1_0 d\alpha \int_0^{\infty} d\rho
\exp\left\{- \rho\left(M^2 + \alpha(1-\alpha)q^2 \right) - \frac{(\theta q)^2}{4\rho}\right\}.
\end{eqnarray}
Here $K_0(z)$ is the modified Bessel function.
The effective four point coupling of the fermions with momentum
transfer $q$ is

\begin{equation}
\lambda_{\rm eff}(q)=
\frac{1}{\frac{1}{\lambda} + \tau_1(q) + \tau_2(q)} 
\label{coupling}
\end{equation}
When we substitute the cut-uff-dependent expression (\ref{gap}) for
$1/\lambda$ into (\ref{coupling}), the UV cutoff dependence does not
cancel.  If $q>1/\theta\Lambda$, the effective coupling $\lambda_{\rm
eff}(q)$ goes to zero as $\Lambda$ is taken to infinity. 

\noindent
{\bf 3.} Let us find UV behaviour of (\ref{coupling}). In the limit
when $q^2 >> 4 M^2$, $q^2 >> 1/(\theta M)^2$ (we always assume that
$q^2<<\Lambda^2$ and $M^2<<\Lambda^2$) we have

\begin{equation}
\tau_1(q)\approx - \frac{1}{8\pi}\ln\frac{\Lambda^2}{q^2} 
\quad {\rm and} \quad
\tau_2(q)\sim e^{-\theta M q}
\end{equation}
Thus

\begin{equation}
\lambda_{\rm eff}(q) \approx 
\frac{1}{\frac{1}{\lambda} - \frac{1}{8\pi}\ln\frac{\Lambda^2}{q^2}} 
=
\frac{8\pi}{\ln\frac{\Lambda^2q^2}{M^4}},
\end{equation}
where $\lambda$ is eliminated using (\ref{gap}).

On the other hand, when $q^2<<1/(\theta^2M^2)$, $q^2 << 4 M^2$ and
$q^2 >> 1/(\theta \Lambda)^2$ we can approximate the above expressions
by

\begin{eqnarray}
\tau_1(q) \approx - \frac{1}{8\pi}\ln\frac{\Lambda^2}{M^2} 
\quad {\rm and} \quad
\tau_2(q) \approx - \frac{1}{8\pi} \ln\left(\theta^2q^2M^2\right)
\end{eqnarray}
Hence

\begin{eqnarray}
\lambda_{\rm eff}(q) \approx \frac{1}{\frac{1}{\lambda} 
- \frac{1}{8\pi}\ln\frac{\Lambda^2}{M^2} 
- \frac{1}{8\pi} \ln\left(\theta^2q^2M^2\right)} =
\frac{8\pi}{\ln\left(\Lambda^2 \theta^2 q^2\right)}.
\end{eqnarray}
For momenta above $q\sim 1/\theta\Lambda$ the last expression depends
on the cutoff and for finite, nonzero momentum it goes to zero as the
cutoff goes to infinity.

  Quite interesting things happen in the double scaling limit when
  $\Lambda\to\infty$ and $\theta\to 0$ so that $\Lambda\theta = C/M$
  with an arbitrary constant $C$. The physical meaning of this limit
  is that one ``regularizes'' the ordinary Gross-Neveu model by a
  non-commutative one at the cutoff scale.  In this limit we can obtain an
  exact expression:

\begin{eqnarray}
\lambda_{\rm eff}(q) =
\frac{4\pi}{\frac12\ln\left(1 + C^2 q^2/M^2\right) + 
\frac{ \sqrt{1+\frac{q^2}{4M^2}} }
{\frac{q}{2M}}\ln\left(\sqrt{ 1+\frac{q^2}{4M^2}}+\frac{q}{2M}\right)}.
\end{eqnarray}
The second term in the denominator has a squareroot cut starting from
$q=2Mi$ in the complex $q$ plane, which corresponds to a pair
production of fermions.  This is the same as what occurs in the
commutative Gross-Neveu model.  What is new and interesting is the
first term.  It has a logariphmyc cut starting from $q=iM/C$. This cut
is absent in the commutative model. It probably corresponds to a
creation of pairs of some non-local solitons present in the
non-commutative theory, which survive the double scaling limit.  We
see that the limits $\Lambda\rightarrow\infty$ and $\theta\rightarrow
0$ do not commute and even in the case when non-commutativity is
relevant at the cutoff scale it still modifies the behaviour of the
theory at any energy scale.

\noindent
{\bf 4.} We have found that if we allow $\Lambda$ to go to infinity
with fixed $\theta$, this particular asymptotically free theory
is trivial.
We argue that this is a generic feature of non-commutative field
theories.  In fact, consider the following general physical arguments.
An excitation with large momentum $p_x$ in a non-commutative theory
has uncertainty in its position along the momentum $\Delta x \sim
1/p_x$ which is very small. Hence, taking into account that $x$ and
$y$ coordinates do not commute, we see that the uncertainty in $y$ is
very big. This mixes IR and UV limits in the sense that IR effects
modify UV limit and vise versa \cite{Susskind}. As we see from our
example, this mixing is generally model independent and generic for
theories with a dynamically generated mass gap.  Taken to its extreme
conclusion it would imply that all non-commutative theories with
asymptotic freedom are trivial and therefore, in particular, the limit
of open string theory in a constant $B$-field is a trivial theory
unless the ultraviolet cutoff $1/\alpha'$ is kept finite.

It is straightforward to generalize our results to the 2+1 dimensional
Gross-Neveu model and to 2+1 and 3+1-dimensional $O(N)$ vector model with
quartic interactions of scalars and {\it space-space} non-commutativity.
There, the conclusions are even more drastic than in the present two
dimensional case. 

This work is supported in part by NSERC of Canada.  E.T.A. was
supported by a NATO Science Fellowship and grants INTAS-97-01-03 and
RFBR 98-02-16575.  We thank Kostya Zarembo for valuable discussions.


\end{document}